\documentclass[10pt, conference,letterpaper]{IEEEtran}
\usepackage[letterpaper, left = 0.68in, right =0.68in, top = 0.75in, bottom = 1.042in]{geometry}

\def\arxivdisclaimer{1}

\def\longVersion{1}

\usepackage[noadjust]{cite}
\usepackage{amsmath}
\usepackage{amssymb}
\usepackage{amsthm}
\usepackage{amsfonts}
\usepackage{graphicx}
\usepackage{textcomp}
\usepackage{mathtools}
\usepackage{adjustbox}
\usepackage{balance}
\usepackage{verbatim}
\usepackage{bm}
\usepackage{tikz}
\usepackage{titlesec}
\usepackage{float}
\usepackage{siunitx}
\usepackage{gensymb}

\usetikzlibrary{arrows, fit, matrix, positioning, shapes, backgrounds, spy}
\usepackage[dvipsnames]{xcolor}
\def\BibTeX{{\rm B\kern-.05em{\sc i\kern-.025em b}\kern-.08em
    T\kern-.1667em\lower.7ex\hbox{E}\kern-.125emX}}

\usepackage{textcase}
\usepackage[tablename=TABLE,font=small,compatibility=false]{caption}
\DeclareCaptionTextFormat{up}{\MakeTextUppercase{#1}}

\usepackage{paralist}
\usepackage{supertabular}    
\usepackage{array} 
\usepackage[linesnumbered,ruled]{algorithm2e}
\usepackage{algorithmic}
\usepackage[acronym]{glossaries}

\usepackage{numprint}     
\usepackage{makecell}   
\usepackage{colortbl}
\usepackage{enumitem}            
\usepackage{longtable} 
\usepackage{wrapfig}
\usepackage{booktabs}
\usepackage{graphics}
\usepackage{pgfplots}
\usepgfplotslibrary{groupplots}
\usepackage{pgfplotstable}
\usepackage{subcaption}
\pgfplotsset{compat=1.18} 
\usepackage{xfakebold}

\newcommand{\fbseries}{\unskip\setBold\aftergroup\unsetBold\aftergroup\ignorespaces}
\makeatletter
\newcommand{\setBoldness}[1]{\def\fake@bold{#1}}
\makeatother

\usepackage{tabu}

\newcommand{\mi}[1]{\mathit{#1}}
\newcommand{\mr}[1]{\mathrm{#1}}
\newcommand{\Tx}{\mathrm{Tx}}
\newcommand{\Rx}{\mathrm{Rx}}

\newtheorem*{remark*}{Remark}

\usepackage[utf8]{inputenc}

\let\oldtabular\tabular
\renewcommand{\tabular}{\small\oldtabular}

\newcommand{\iec}{i.\,e., }

\newcolumntype{?}{!{\vrule width 1pt}}
\definecolor{mittelblau}{RGB}{0, 126, 198}
\definecolor{violettblau}{cmyk}{0.9, 0.6, 0, 0}
\definecolor{rot}{RGB}{238, 28 35}
\definecolor{apfelgruen}{RGB}{140, 198, 62}
\definecolor{gelb}{RGB}{1, 221, 0}
\definecolor{orange}{RGB}{244, 111, 33}
\definecolor{pink}{RGB}{237, 0, 140}
\definecolor{lila}{RGB}{128, 10, 145}
\definecolor{hellgrau}{RGB}{224, 224, 224}
\definecolor{mittelgrau}{RGB}{128, 128, 128}
\definecolor{dunkelgrau}{RGB}{80,80,80}
\definecolor{anthrazit}{RGB}{19, 31, 31}
\definecolor{darkgreen}{RGB}{0.125,0.5,0.169}
\definecolor{ahmedyellow}{RGB}{204,153,0}

\newcommand\blfootnote[1]{%
  \begingroup
  \renewcommand\thefootnote{}\footnote{#1}%
  \addtocounter{footnote}{-1}%
  \endgroup
}

\if\longVersion0
\titlespacing*{\section}
{0pt}{*1.5}{*0.75}
\fi

\begin{document}

\title{Reliable UAV Detection with ISAC}

\author{
    \IEEEauthorblockN{
        Stephan Saur,
        Mark Doll,
        Artjom Grudnitsky,
        Silvio Mandelli,
        Lucas Giroto,
        Marcus Henninger,
        Thorsten Wild
        }

	\IEEEauthorblockA{Nokia Bell Labs Stuttgart, Germany
        \\
	E-mail:\{stephan.saur, mark.doll, artjom.grudnitsky, silvio.mandelli, lucas.giroto, marcus.henninger, thorsten.wild\}\\@nokia-bell-labs.com}}

\maketitle

\newacronym{1D}{1D}{one-dimensional}
\newacronym{2D}{2D}{two-dimensional}
\newacronym{3GPP}{3GPP}{3rd Generation Partnership Project}
\newacronym{6G}{6G}{sixth generation}
\newacronym{adc}{ADC}{analog-to-digital converter}
\newacronym{awgn}{AWGN}{additive white Gaussian noise}
\newacronym{cacfar}{CA-CFAR}{cell averaging constant false alarm rate}
\newacronym{crap}{CRAP}{clutter removal with acquisitions under phase noise}
\newacronym{csi}{CSI}{channel state information}
\newacronym{cp}{CP}{cyclic prefix}
\newacronym{dft}{DFT}{discrete Fourier transform}
\newacronym{dl}{DL}{downlink}
\newacronym{eca-c}{ECA-C}{extensive cancellation algorithm by subcarrier}
\newacronym{fdr}{FDR}{false discovery rate}
\newacronym{fft}{FFT}{fast Fourier transform}
\newacronym{gnb}{gNB}{next-generation Node B}
\newacronym{gnss}{GNSS}{global navigation satellite system}
\newacronym{hpbw}{HPBW}{half-power beam width}
\newacronym{idft}{IDFT}{inverse discrete Fourier transform}
\newacronym{ifft}{iFFT}{inverse fast Fourier transform}
\newacronym{iip3}{IIP3}{3rd order intermodulation input intercept point}
\newacronym{isac}{ISAC}{integrated sensing and communication}
\newacronym{kf}{KF}{Kalman filter}
\newacronym{ofdm}{OFDM}{orthogonal frequency-division multiplexing}
\newacronym{oip3}{OIP3}{3rd order intermodulation output intercept point}
\newacronym{omp}{OMP}{orthogonal matching pursuit}
\newacronym{psd}{PSD}{power spectral density}
\newacronym{psf}{PSF}{point spread function}
\newacronym{poc}{PoC}{proof-of-concept}
\newacronym{rcs}{RCS}{radar cross section}
\newacronym{rf}{RF}{radio frequency}
\newacronym{rmse}{RMSE}{root-mean-square error}
\newacronym{ru}{RU}{radio unit}
\newacronym{rx}{RX}{receiver}
\newacronym{snr}{SNR}{signal-to-noise ratio}
\newacronym{sinr}{SINR}{signal-to-interference-plus-noise ratio}
\newacronym{spu}{SPU}{sensing processing unit}
\newacronym{sqnr}{SQNR}{signal-to-quantization noise ratio}
\newacronym{tdd}{TDD}{time division duplex}
\newacronym{tx}{TX}{transmitter}
\newacronym{uav}{UAV}{unmanned aerial vehicle}
\newacronym{ul}{UL}{uplink}
\begin{abstract}

\Acrfull{uav} detection is one prominent use case of \acrfull{isac} systems in 5G-Advanced and future 6G networks. In this paper, we present experimental results for the  detection of a small \acrshort{uav} using unmodified commercial 5G hardware for mono-static \acrfull{ofdm} radar and compare them with the expected performance based on models for link budget and hardware impairments. We show that reliable detection with sub-meter accuracy is still possible in 500 meters distance in a challenging radio environment rich of strong clutter.
\end{abstract}

\if\arxivdisclaimer1
\blfootnote{This work has been submitted to the IEEE for possible publication. Copyright may be transferred without notice, after which this version may no longer be accessible.}
\else
\vspace{0.2cm}
\fi

\if\longVersion1
\else
\vspace{-5mm}
\fi

\glsresetall
\section{Introduction}\label{sec:intro}
Sensing capabilities will complement future cellular 5G-Advanced and 6G mobile radio networks beyond pure communication services and thus enable an entirely new class of use cases \cite{ghosh2025unified}. This is possible thanks to \gls{ofdm}-based radar~\cite{braun2014ofdm} operated by cellular base stations and terminals. Therefore, \gls{isac} systems allow to detect objects' distance (\textit{range} in radar terminology), direction, and speed, regardless of whether they are connected to the network.

One prominent use case of \gls{isac} systems that is closely aligned with \gls{3GPP}'s objectives \cite{azim2025rcs} is the detection and tracking of \glspl{uav} in the lower airspace for the sake of monitoring low altitude economy, protection of critical infrastructure, intrusion detection, border surveillance and defense needs. This use case is particularly relevant as non-cooperative \glspl{uav} can be misused to enable jamming and spoofing attacks to disrupt \gls{gnss} navigation and cellular networks, while also performing unauthorized environment sensing with sensors such as radars or cameras \cite{beuster2025swarm}. While \glspl{uav} can alternatively be detected via their transmitted radio signals~\cite{thomae2025distributed}, \gls{isac} also enables detection of stealth \glspl{uav} that only transmit sporadically if at all. Further advantages of using \gls{isac} for \gls{uav} detection include the availability and ubiquitous coverage of a nation-wide cellular radio network, such that in the ideal case no additional equipment would need to be deployed in the field.

In \cite{mandelli2023survey}, the theoretical performance potential of \gls{isac} has been derived depending on system limitations. Building upon this, the contributions of this paper are (i) setup of an \gls{isac} \gls{poc} in a realistic outdoor scenario, (ii) derivation of a link budget model considering the impairments of commercial 5G hardware, and (iii) experimental validation of the theoretical limitations. The main challenge addressed is detecting \glspl{uav} with small \gls{rcs} in the presence of clutter, \iec strong unwanted signal components reflected by other static or dynamic objects. 
\section{Proof of Concept}\label{sec:poc_intro}

\begin{figure}
\begin{minipage}[t]{0.32\columnwidth}
      \vspace{0pt}
      \begin{subfigure}{\textwidth}
      \includegraphics[width=\columnwidth]{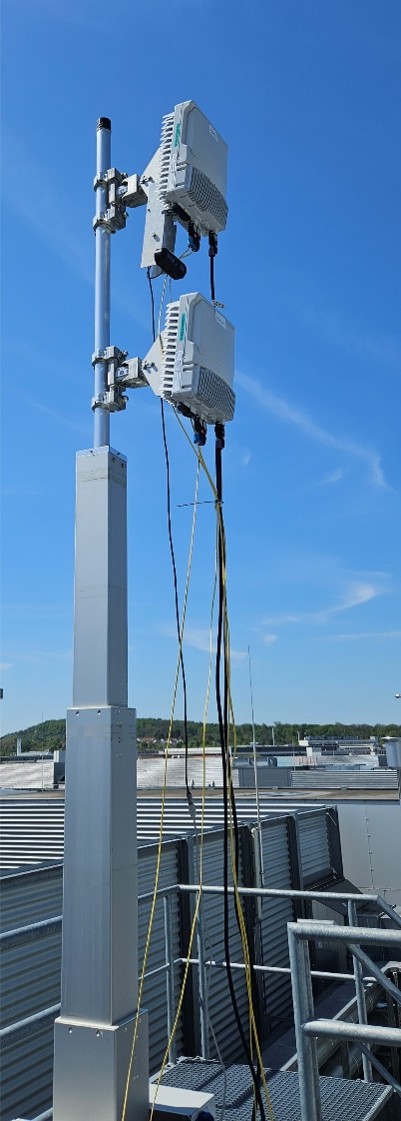}
      \caption{RU deployment.}
      \label{fig:tc2-rus}
      \end{subfigure}
\end{minipage}%
  \hfill
\begin{minipage}[t]{0.65\columnwidth}
    \vspace{0pt}
    \begin{subfigure}{\textwidth}
    \centering
    \includegraphics[width=0.8\columnwidth]{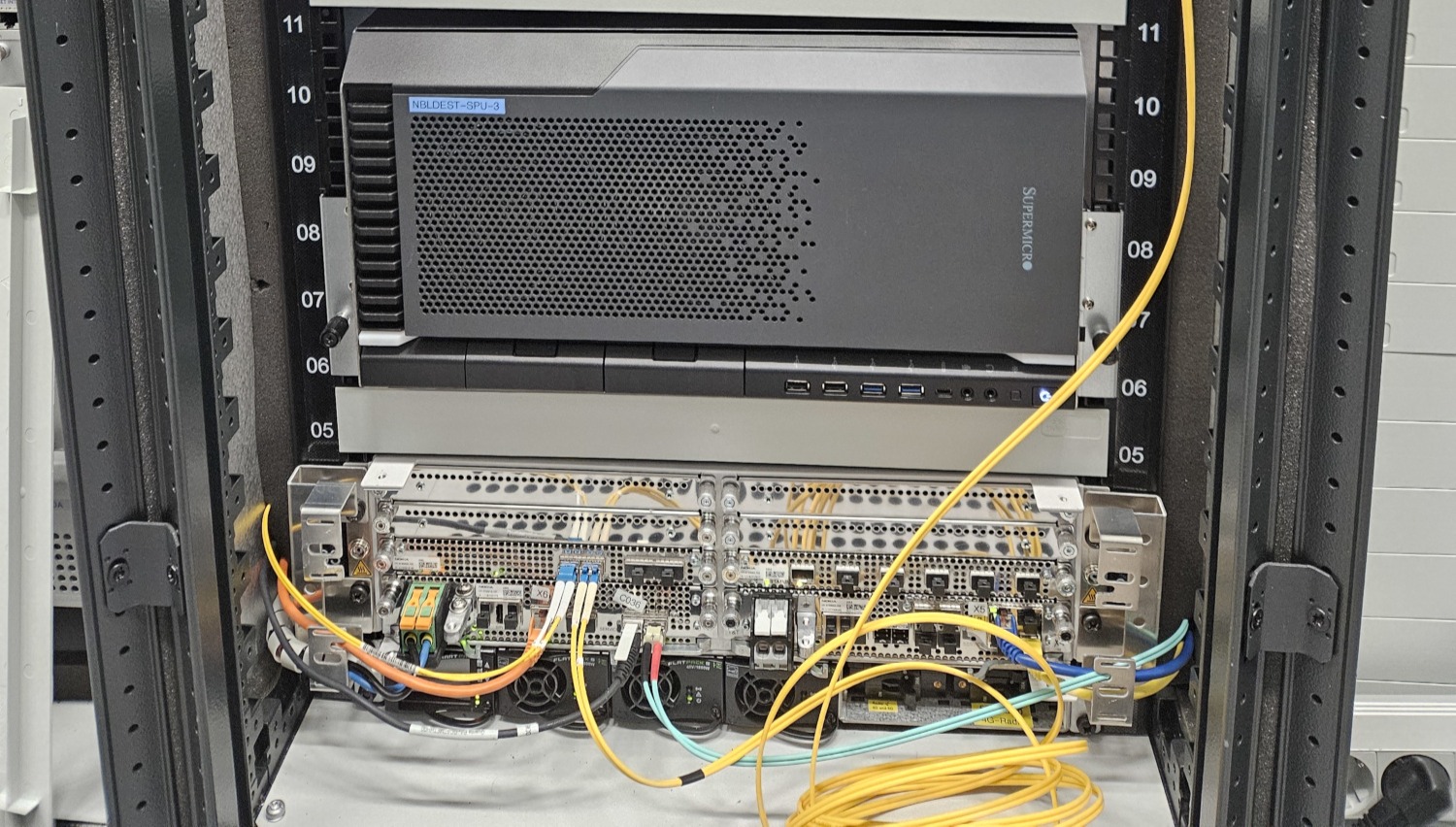}
    \caption{SPU server.}
    \label{fig:spu-server}
    \end{subfigure}

    \vspace{4em}

    \begin{subfigure}{\textwidth}
    \centering
    \includegraphics[width=0.8\columnwidth, angle =0]{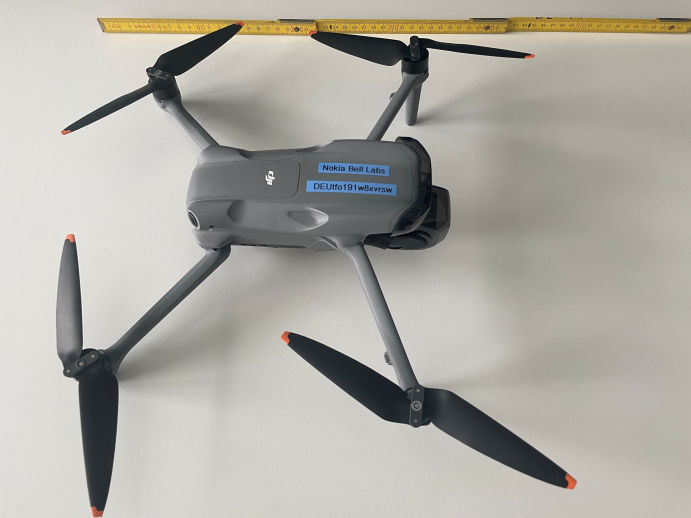}
    \caption{DJI Air 3 \gls{uav}.}
    \label{fig:drone}
    \end{subfigure}
\end{minipage}
  \caption{\gls{isac} \gls{poc} deployment and \gls{uav} used in sensing experiments.}
  \label{fig:poc-and-drone}
  \vspace{-0.25cm}
\end{figure}

For the presented \gls{uav} detection experiments we used throughout this paper an \gls{isac} \gls{poc} \cite{wild2023poc} consisting of commercial 5G mmWave \glspl{ru} for the transmission of radio frames illuminating the environment and the reception of the reflections.
The \glspl{ru} are connected to the \gls{spu}, an off-the-shelf GPU server (Fig.~\ref{fig:spu-server}) running custom software for controling and configuring the \glspl{ru}, as well as performing sensing processing, which is based on \gls{ofdm}-radar~\cite{braun2014ofdm}.

\begin{figure}
  \begin{center} 
      \includegraphics[width=1.0\linewidth]{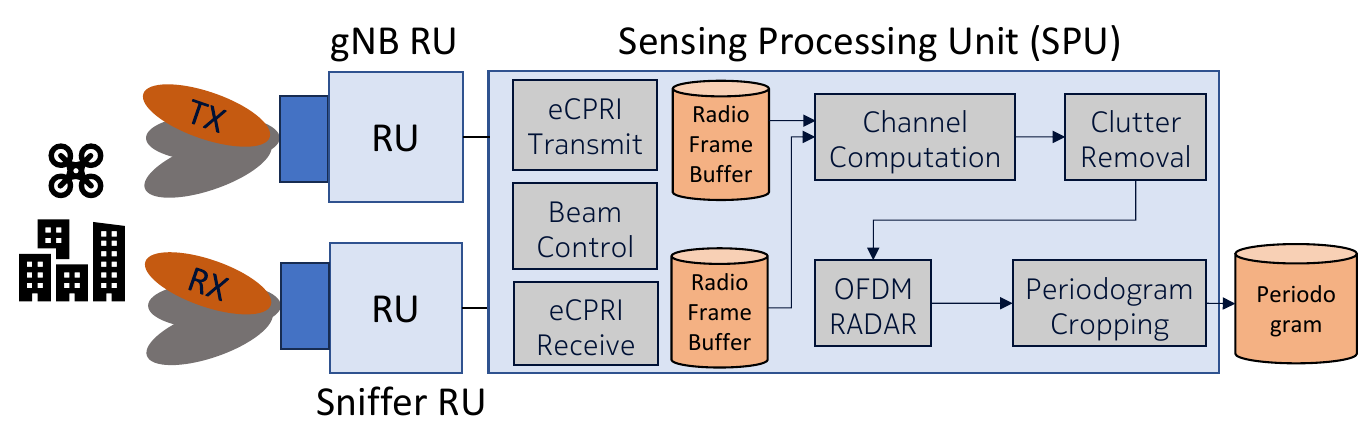}
      \caption{\gls{isac} PoC processing flow diagram.}
      \label{fig:full-system-overview}
  \end{center}
  \vspace{-0.25cm}
\end{figure}

Fig.~\ref{fig:full-system-overview} shows the block diagram of the \gls{poc} processing flow.
The eCPRI Transmit and Beam Control modules in the SPU transmit frequency-domain I/Q OFDM radio frames, beam control and \gls{tdd} configuration information to the \gls{gnb} RU, which is used to illuminate the environment.
The Sniffer RU is switched to \gls{ul} mode, while the gNB RU is in \gls{dl} mode, to allow it receiving the environmental reflections and send them via the fronthaul link to the SPU eCPRI Receive module as frequency-domain I/Q data.
Once a full radio frame of reflections is received, the SPU computes the time-frequency channel by element-wise division of the received radio frame by the transmitted radio frame.
Optional clutter removal can be performed to suppress static clutter such as buildings.
The channel is then processed using batched \gls{fft} followed by batched iFFT operations to generate a range-Doppler periodogram for each 10~ms radio frame.
These periodograms are then used for the \gls{uav} detection analysis.

\begin{figure}
    \centering
    \includegraphics[width=0.8\linewidth]{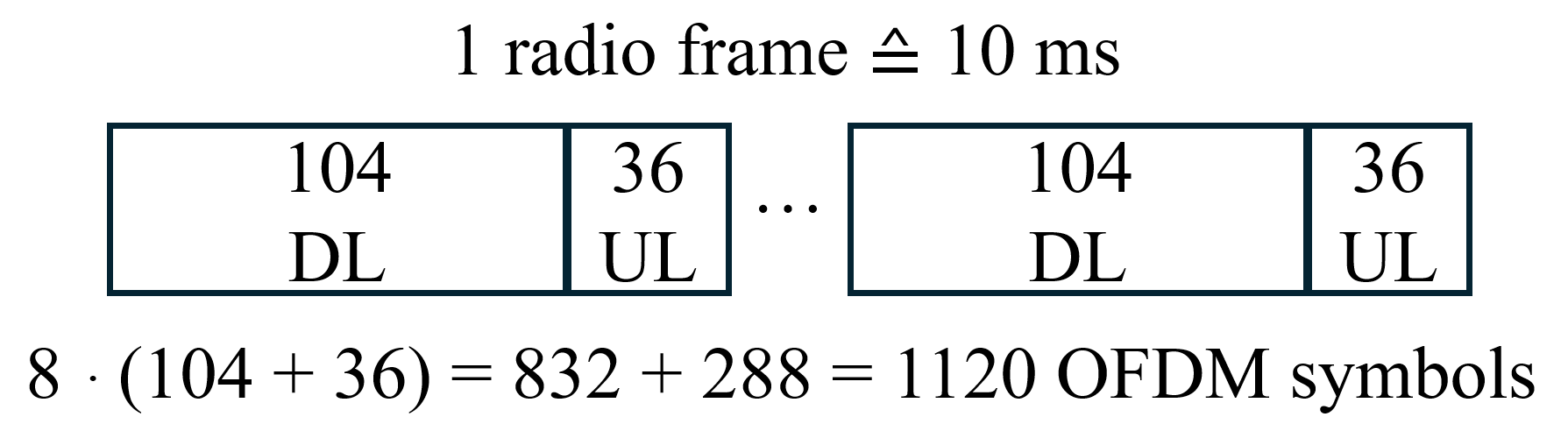}
    \caption{TDD frame structure}
    \label{fig:frame-config}
\end{figure}

For the experiments in this paper, we use a deployment on the rooftop of a building (Fig.~\ref{fig:tc2-rus}), with the gNB RU and Sniffer RU co-located in a quasi-monostatic setup at a height of 27~m above street level.
The deployment covers an inner-city valley with clutter consisting of buildings, traffic and hills.
The \gls{poc} is configured for 200~MHz bandwidth in 5G FR2 $\mu=3$ numerology \cite{3gpp_38211}.
The gNB RU is using a TDD frame structure with 832 \gls{dl} \gls{ofdm} symbols distributed as shown in Fig.~\ref{fig:frame-config} and 1584 subcarriers for sensing, \iec $\approx~75\%$ of all radio resources.
The \glspl{ru} are set up not to radiate above the horizon, thus clutter is present in all measurements.
Further PoC configuration parameters are listed in Table~\ref{tab:linkbudgetparams}.

\section{Link Budget Model}\label{sec:lb_model}

For the theoretical derivation of the maximum distance $r_\mr{max}$ at which the \gls{uav} should be detectable, we use the following link budget model~\cite{mandelli2023survey}

\begin{align}
    r^* = \left(
    \smash{\overbrace{
      \underbrace{
        P_\Tx\frac{M}{\Delta f} \phantom{\rule[-2.5ex]{1pt}{6.0ex}}
        }_{\mr{Tx~energy}}
      \underbrace{ \phantom{\rule[-2.5ex]{1pt}{6.0ex}}
        G_\mr{Tx}\frac{1}{(4\pi)^3}\sigma\frac{c^2}{f_\mr{c}^2}G_\mr{Rx}}_{\mr{coupling~loss}\smash{^{-1}}\ @1m}
      \frac{1}{S_\mr{N+I}}}^{\mi{SNR}_0\ =\ \mr{signal\text-to\text-noise~ratio~at~unit~range}}
    }
    \frac{1}{\gamma_\mr{min}}\right)^{1/4}. \label{eq:bestrange}
\end{align}
\par\vspace*{15pt}
In this equation, $r^*$ is the achievable distance (with receive window shifted to match propagation delay), $P_\Tx$ is the total radiated power of the transmitter, $M$ the number of \gls{ofdm} symbols used for sensing, $\Delta f$ the subcarrier spacing, whose inverse equals the useful \gls{ofdm} symbol duration, $G_\Tx$ and $G_\Rx$ the antenna gains of transmitter and receiver, respectively, $\sigma$ the \gls{rcs} of the reflective object, in our case the \gls{uav}, $c$ the speed of light (in air), $f_\mr{c}$ the carrier frequency, $S_\mr{N+I}$ the noise and interference \gls{psd} measured at the receiver input, and $\gamma_\mr{min}=17~\text{dB}$ (after considering all gains) the minimum \gls{sinr} to achieve robust performance~\cite{johnson2008gamma}. For the \gls{psd} of noise plus interference $S_\mr{N+I}$, we consider thermal noise $N_\mr{0,Rx}$ of the receiver and 3rd order intermodulations from transmitter $S_\mr{3,Tx}$ and receiver $S_\mr{3,Rx}$, respectively 
\begin{gather}
S_\mr{N+I}=N_{0,\mr{Rx}}+S_\mr{3,Tx}+S_\mr{3,Rx}, \\
N_\mr{0,Rx} = N_0 \mi{F}_\Rx, \\
S_\mr{3,Tx} = \frac{1}{N\Delta f} \frac{P_\Rx}{(\mi{OIP3}_\Tx/P_\mr{PAout})^2}, \\
S_\mr{3,Rx} = \frac{1}{N\Delta f} \frac{P_\Rx}{(\mi{IIP3}_\Rx/P_\mr{LNAin})^2}.
\end{gather}
$N_0=k_\mr{B} T$ is the thermal noise \gls{psd} at absolute temperature $T$ with Boltzmann constant $k_\mr{B}$ and $\mi{F}_\Rx$ the receiver noise figure.
The power of the $n$-th order intermodulation product is $P_n=P_\mr{out}/(\mi{IIPn}/P_\mr{in})^{n-1}$~\cite{korpi2014}.  Division by the useful carrier bandwidth $N\Delta f$ with number of used subcarriers $N$ converts power $P_i$ to \gls{psd} $S_i$. Only the strongest 3rd order intermodulation is relevant in practice and thus only the 3rd order intermodulation input and output intercept points $\mi{IIP3}_\Rx$, respectively, $\mi{OIP3}_\Tx$ are specified in datasheets \cite{admv4801} and considered here. Note that input and output power or \gls{psd} as well as input and output intercept points are related via the amplifier gain $G$ as $P_\mr{out}=G\,P_\mr{in}$ and $\mi{OIPn}=G\,\mi{IIPn}$.
Furthermore,
\begin{gather}
P_\mr{PAout} = \frac{P_\Tx}{\mi{N}_\Tx}, \\
P_\mr{LNAin} = P_\Tx\left(\frac{1}{\mi{C}_\mr{total}\mi{N}_\Rx} + \frac{1}{I}\right), \\
P_\Rx = P_\Tx\left(\frac{1}{\mi{C}_\mr{total}} + \frac{1}{I}\right).
\end{gather}
$P_\mr{PAout}$ is the output power of a single of the $\mi{N}_\Tx$ radiators overall in the transmitter phased array, and $P_\mr{LNAin}$ is the input power per single radiator in the receiver phased array of $\mi{N}_\Rx$ radiators.
The total receive power $P_\Rx$ consists of the received sensing signal  $P_\Tx/\mi{C}_\mr{total}$ and self-interference $P_\Tx/I$, which differ in that only the former includes the antenna gains $G_\Tx$ and $G_\Rx$.
$\mi{C}_\mr{total}$ is the total coupling loss over all clutter and targets, and $I$ the average isolation between a pair of Tx and Rx radiators.\par
For a receive window of duration $T_\mr{sym}=1/\Delta f$, which is aligned with the end of the transmitted \gls{ofdm} symbol of duration $T_0$, the propagation delay of a target at a distance beyond $r_\mr{CP}=T_\mr{CP}c/2$ exceeds the \gls{cp} duration $T_\mr{CP}=l_\mr{CP}/\Delta f$ and a shrinking fraction of the reflected signal from the target falls into the receive window, causing the detected target signal's magnitude to decrease linearly with the propagation delay until it reaches zero at the end of the receive window for targets at distance $r_\mr{limit}$.
The reason is that both the fraction of received transmit energy and captured sensing signal sequence (in time) and thus processing gain are proportional to the overlap of sensing signal with receive window, resulting in quadratic decrease in power respectively linear decrease in magnitude.
The cyclic prefix provides protection against the strongest inter-symbol-interference from the previous symbol(s) directly from the transmitter and from nearby and thus strongest targets up to $r_\mr{cp}$ and is therefore ignored here.
Delaying the receive window by $T_\mr{rx}=l_\mr{rx}/\Delta f$ with sensing receive window offset $l_\mr{rx}$, ideally according to the target signal's expected propagation delay, extends the maximum distance by up to $r_\mr{rx}=T_\mr{rx}c/2$, but at the same time causes signals from nearby targets to arrive partly prior to the begin of the receive window, similarly causing a linear decrease in magnitude until not detecting any targets anymore at distances below $r_\mr{low}$. The resulting reduced maximum distance is
\begin{align}
\arraycolsep=0pt
r_\mr{max}=\left\{
\begin{array}{cl}
    a+\sqrt{a^2-2a(r_\mr{rx}-r_\mr{sym})},\, & r^* \in [r^*_\mr{min},r_\mr{rx}] \\
    r^*,\,                                   & r^* \in [r_\mr{rx},r_\mr{rx}+r_\mr{CP}] \\
    -a+\sqrt{a^2+2a(r_\mr{rx}+r_0)},\,       & r^* > r_\mr{rx}+r_\mr{CP} \\
\end{array}\right.,
\label{eq:maxrange}
\end{align}
where $a={r^*}^2/2r_\mr{sym}$ captures the impact of the limited receive window of duration $T_\mr{sym}$, $r_\mr{sym}=T_\mr{sym}c/2$ and $r_0=T_0c/2$ being the target ranges where the propagation delay equals the duration of the receive window and the transmitted \gls{ofdm} symbol including \gls{cp} $T_0=(1+l_\mr{CP})/\Delta f$, respectively, and $l_\mr{CP}=1/14$ is the \gls{cp} length expressed as fraction of the useful \gls{ofdm} symbol duration $T_\mr{sym}=1/\Delta f$ \cite{3gpp_38211}.
Note that for a shifted receive window $T_\mr{rx}>0$, an equal guard time is assumed where the transmitter stays off.
No target closer than $r_\mr{low}=r_\mr{rx}-r_\mr{sym}$ or beyond $r_\mr{limit}=r_\mr{rx}+r_0$ can be detected, where all of the reflected signal arrives outside of the receive window.
Weak targets that would be detected with perfect receive window alignment up to $r^*_\mr{min}=2\sqrt{r_\mr{sym}(r_\mr{rx}-r_\mr{sym})}$ are detected only up to $2r_\mr{low}$, even weaker targets cannot be detected anymore.
In our experiments the receive window was not delayed, \iec $T_\mr{rx}=0$.

All system and hardware related parameters of our \gls{poc} used for the link budget calculation are given in Table~\ref{tab:linkbudgetparams}.

\begin{table}[t]
    \caption{System and hardware parameters. \label{tab:linkbudgetparams}}
    \centering
    \begin{tabu}{|l|l|l|}
         \hline
         \textbf{Symbol} & \textbf{Parameter}  & \textbf{Value} \\
         \Xhline{3\arrayrulewidth}
         $f_\mr{c}$ & Carrier frequency & 27.6~GHz \\
         \hline         
         $\Delta f$ & Subcarrier spacing & 120~kHz \\
         \hline
         $l_\mr{CP}$ & Average cyclic prefix length & 1/14 \\ 
         \hline
         $T_\mr{rx}$ & Sensing Rx time offset (vs.\ Tx) & 0~s\\
         \hline
         $N$ & Number of used subcarriers & 1584 \\
         \hline
         $M$ & Number of used \gls{ofdm} symbols & 832 \\
         \hline
         $P_\Tx$ & Total radiated power & 28.1~dBm \\
         \hline
         $\mi{OIP3}_\Tx$ & Tx 3rd order output intercept point & 23.1~dBm \\
         \hline
         $\mi{N}_\Tx$=$\mi{N}_\Rx$ & Array factor of Tx and Rx & 96 \\
         \hline
         $G_\Tx=G_\Rx$ & Antenna gain of Tx and Rx & 23.4~dB \\
         \hline
         $I$ & Tx-Rx isolation (single radiator) & 103~dB \\
         \hline
         $\mi{C}_\mr{total}$ & Sum coupling loss incl. clutter & 85~dB \\
         \hline
         $\mi{IIP3}_\Rx$ & Rx 3rd order input intercept point & $-13.3$~dBm \\
         \hline
         $N_0=k_\mr{B}T$ & Thermal noise spectral density & $-174$ dBm/Hz \\
         \hline         
         $\mi{F}_\Rx$ & Receiver noise figure & 5.0~dB \\
         \hline
\if10
         $A_\Rx$ & Receiver total amplifier gain & 24.8~dB \\
         \hline
         $P_\mr{0dBFS}$ & \gls{adc} 0~dB full scale input power & 1~dBm \\
         \hline
         $N_\mr{bit}$ & \glspl{adc} resolution & 14~bit \\
         \hline
         $N_\mr{ADC}$ & Number of \glspl{adc} & 1 \\
         \hline      
\fi
    \end{tabu}
\end{table}
\section{Experiments}\label{sec:experiments}

For our \gls{uav} detection experiments with the \gls{isac} PoC, we used the DJI Air~3 shown in Fig.~\ref{fig:drone}. The diagonal distance between the ends of the unfolded propellers is 56~cm. Based on the comparison of different \gls{uav} types \cite{semkin2020rcs}, we have assessed its mean \gls{rcs} as $\sigma= -17~\text{dBsm}$. 

\begin{figure}
    \centering
    \includegraphics[height=6.8cm]{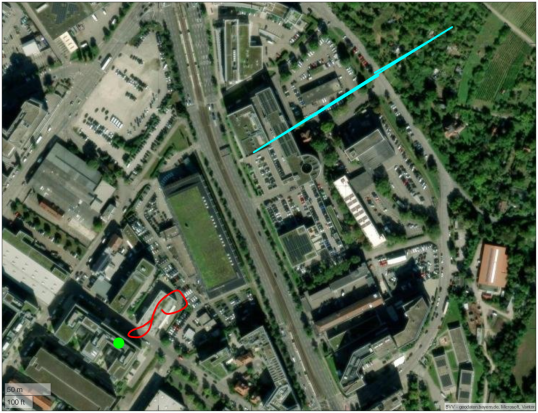}%
    \llap{\raisebox{0cm}{
      \includegraphics[height=3.5cm]{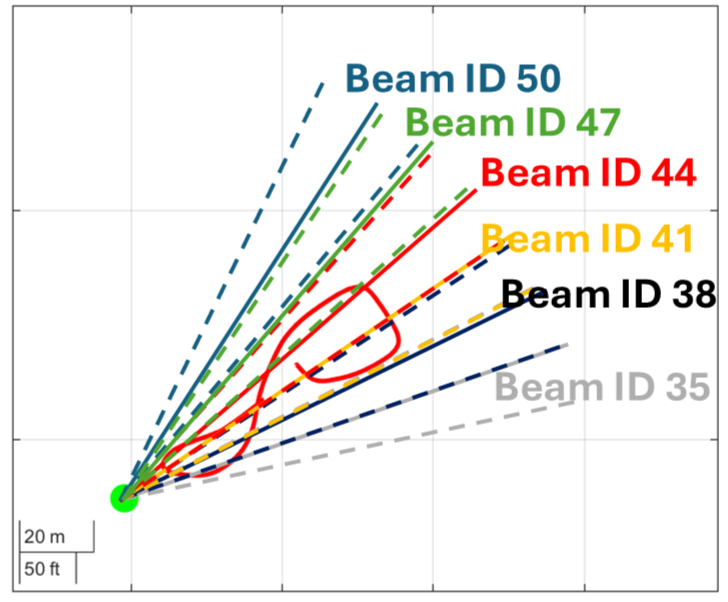}%
    }}
    
    \caption{Investigated flight routes (red and light blue trajectory), position of \glspl{ru} (green dot); zoomed in: boresight direction (solid) and \gls{hpbw} (dashed) of each beam ID.}

    \label{fig:maproutes}
    \vspace{-0.25cm}
  \end{figure}

In a first experiment, we navigated the \gls{uav} on a random flight route, shown in Fig.~\ref{fig:maproutes} as the red trajectory, with a maximum distance of 90~m to the \glspl{ru}, marked in green. The \gls{uav} was always in line-of-sight and at the same height as the base station antennas. During the measurement, we applied a sweep over six beam directions. The top view of the boresight directions of the beams are shown as straight lines in the zoom in Fig.~\ref{fig:maproutes}. The azimuth \gls{hpbw} is approximately $14\degree$, illustrated as dashed lines. The six beams cover the flight route either fully (ID 41), partially (IDs 35, 38, 44, and 47) or just barely (ID 50). The sweep duration is $300~\text{ms}$, \iec $50~\text{ms}$ per beam direction. \par

We did a second experiment aiming to determine the maximum detectable distance, also illustrated in Fig.~\ref{fig:maproutes} as the light blue trajectory. This environment is characterized by several big office and manufacturing buildings from where we receive strong static reflections. The \gls{uav} was navigated from a central position on a straight route approaching, departing and again approaching the \glspl{ru}.
Due to the longer distance of 250 to 500 meters, we operated the system with one single fixed beam pointing towards the flight route. Extending the route to even longer distances was not possible due to the topology, since rising up the hillside the \gls{uav} leaves the cone of the beam in vertical direction as the elevation \gls{hpbw} is only $6.4\degree$, and pointing the beam upwards was prohibited due to regulatory frequency constraints.\par
The target becomes visible in the periodogram as a correlation peak at a certain delay and Doppler, or, equivalently, range and radial component of the velocity. In order to improve the reliability of the detection, we reduced the impact of clutter before the transformation to the delay-Doppler domain. Two different algorithms have been applied and compared. \Gls{crap} has been discussed in \cite{henninger2024crap2}. The second method based on the \gls{eca-c} \cite{zhao2012multipath} suppresses signal components originating from static reflectors. This is particularly important when the overall signal strength of static clutter is orders of magnitude higher than the desired signal of a small \gls{uav} at the long distances in the second experiment. 100 subsequent radio frames (1 second) per beam direction immediately before the recorded flight have been used for clutter acquisition. As the hovering \gls{uav} was already in the scene during this time, its contribution in the periodogram is suppressed as well. The \gls{uav} becomes detectable only after it starts moving away from its original location. During our experiments we have observed that the stored clutter information becomes outdated after a few minutes, leading to stronger remaining clutter in the periodogram and overall worse detection performance. Hence, clutter acquisition must be repeated frequently in the background. For peak detection, the \gls{cacfar} algorithm is used. \par
In each 10~ms radio frame all 832 DL OFDM symbols have been used for sensing because the intention of our experiment was to find out the maximum distance at which the \gls{uav} can still be detected. Such a sensing-only configuration may be applied in exceptional cases. However, in general, communication and sensing services in an \gls{isac} system will share the available radio resources. Thus, when decimating the radio resources for sensing in time and/or frequency, the processing gain and the \gls{snr} will shrink, and the detection performance is reduced accordingly. Thus, in the case that only a small fraction of the radio resources can be allocated for sensing in average, it is preferable to accumulate these resources in one single radio frame and to reserve other frames for communications only in order to maintain the detection capability at the cost of slightly reduced update rate and increased latency of the sensing service. \par 
The drawback of the \gls{tdd} frame configuration shown in Fig.~\ref{fig:frame-config} is that the acquisition holes due to the empty uplink symbols cause undesired impulsive sidelobes in the radar point spread function in the Doppler domain. However, these replicas can be identified and removed from the list of detected peaks as the wrong speed values do not match with the distance the \gls{uav} moves over time. Additionally, the peak confirmation algorithm proposed in \cite{henninger2025tdd} has been applied for the first experiment. \par
In our experiments, a new target is added to a list of valid targets if a consistent movement of a detected peak is observed during a period of 10 radio frames (100~ms). If not detected for 12 consecutive frames, a target is removed from the list. With this setting, two subsequent beam directions can be bridged within a sweep with which the \gls{uav} is not covered. 
\section{Results}\label{sec:results}

\begin{figure}
    \centering
    \includegraphics[width=0.95\linewidth]{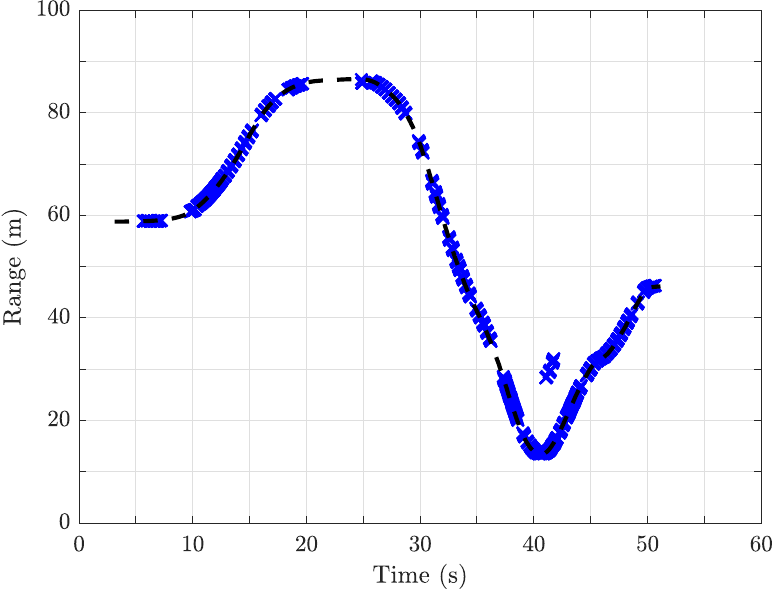}
    \caption{Measured target range from \gls{isac} ({\color[rgb]{0,0,1}{\fbseries$\times$}}) vs.\ recorded \gls{gnss} coordinates of the \gls{uav} ({\color[rgb]{0,0,0}\rule[0.5ex]{0.3em}{.7pt}\hspace{0.3em}\rule[0.5ex]{0.3em}{.7pt}}) for the first experiment.}
    \label{fig:exprange_close}
\end{figure}

The challenge of the first experiment shown as the red trajectory in Fig.~\ref{fig:maproutes} is to continuously detect the \gls{uav} throughout the beam sweep. Fig.~\ref{fig:exprange_close} shows the measured distance of detected targets during the flight. Each detection is represented by a blue cross. The black dashed line is the distance of the \gls{uav} derived from GNSS coordinates that have been recorded by the \gls{uav} itself and acts as ground truth for the evaluation. Time and distance biases between \gls{isac} measurements and GNSS have been removed retrospectively, thus emulating a perfectly synchronized and calibrated system. It has turned out that \gls{crap} combined with the removal of impulsive sidelobes \cite{henninger2025tdd} provides the overall best result with respect to detections. We can see that most of the detections coincide with the flight route of the \gls{uav}. After about 40 seconds, a second target has been identified as valid as well. As can be seen, we could not detect the \gls{uav} in each radio frame of the flight. At the beginning of the flight, the \gls{uav} is still part of the acquired clutter information and removed from the radar image. Moreover, we have observed a fluctuation of the peak power in the recorded measurements during the flight. The applied settings and the achieved range  accuracy of successful detections are summarized in Table~\ref{tab:accuracy}.\par

\begin{figure}
    \centering
    \includegraphics[width=0.95\linewidth]{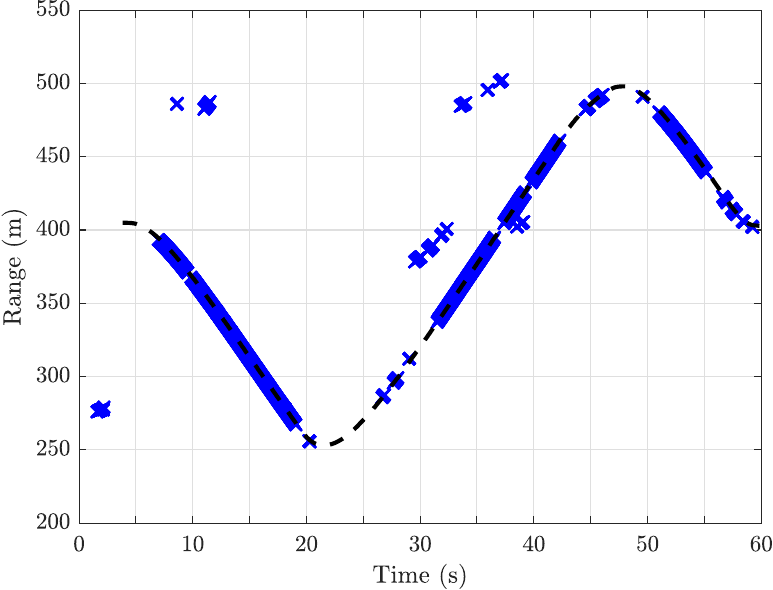}
    \caption{Measured target range from \gls{isac} ({\color[rgb]{0,0,1}{\fbseries$\times$}}) vs.\ recorded \gls{gnss} coordinates of the \gls{uav} ({\color[rgb]{0,0,0}\rule[0.5ex]{0.3em}{.7pt}\hspace{0.3em}\rule[0.5ex]{0.3em}{.7pt}}) for the second experiment.}
    \label{fig:exprange_far}
\end{figure}

The light blue trajectory in Fig.~\ref{fig:maproutes} illustrates the second experiment. The results are shown in Fig.~\ref{fig:exprange_far}. The challenge is to detect the \gls{uav} in spite of the long distance to the \glspl{ru}. In this scenario, \gls{eca-c} is the more reliable method for clutter removal. Reason is that static clutter and its impulsive sidelobes are the predominant components in the reflected signal and must be suppressed to allow for the detection of the much weaker target. In addition, in order to further improve detection rates for the long distance,
the false alarm rate of the \gls{cacfar} detector was increased from $10^{-6}$ to $10^{-4}$. The drawback of ECA-C is that the desired target is suppressed as well while moving slowly, causing missed detections around return points of the route, where the \gls{uav} slowed down before accelerating again into the opposite direction. We also observed that remaining signal components reflected at a big building in 250--300 meters distance lead to a reduced detection probability while the \gls{uav} is flying over it. \par

\begin{figure}
    \centering
    \includegraphics[width=0.88\linewidth]{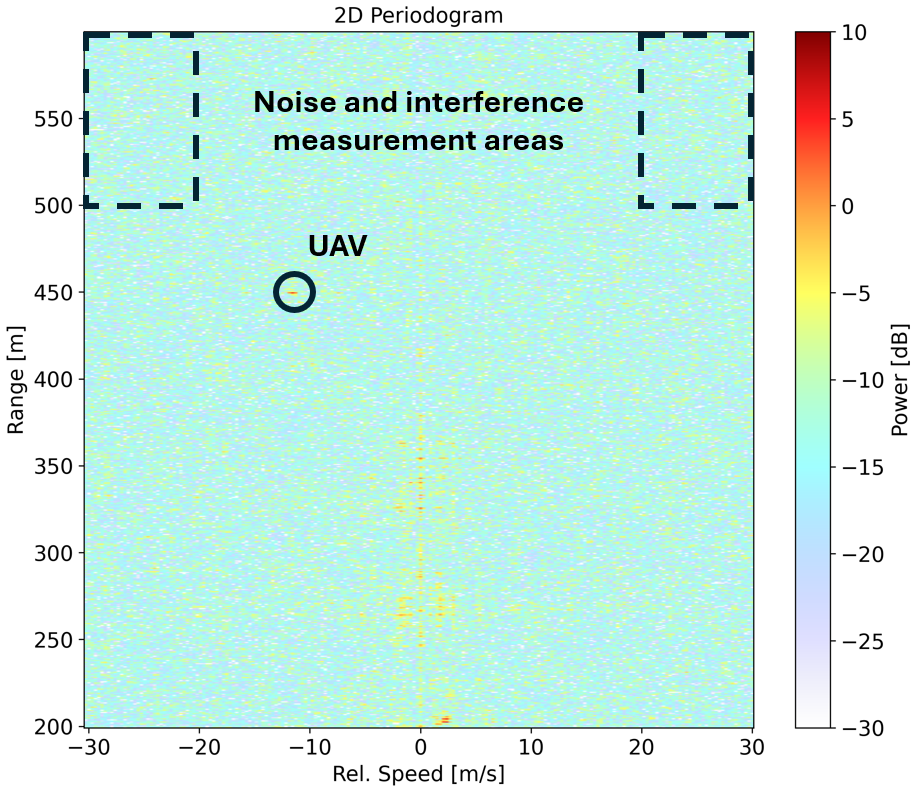}
    \caption{Periodogram showing the detected \gls{uav} in the black circle and the areas used for the assessment of the noise-plus-interference-power within the black dotted lines.}
    \label{fig:per_example}
\end{figure}

For the second experiment we have also measured the \gls{sinr} of all detected and validated peaks in the periodogram. The noise-plus-interference-power consists of the contributions introduced in the link budget model in Section~\ref{sec:lb_model} with power densities $N_\mr{0,Rx}$, $S_\mr{3,Tx}$ and $S_\mr{3,Rx}$, respectively. We have assessed the interference-plus-noise power as the median of the power values in the periodogram which are far away in speed and range from the desired target under the simplified assumption that the power components are equally distributed throughout the periodogram. However, it still also includes residual contributions from undesired clutter objects and their impulsive sidelobes. Fig.~\ref{fig:per_example} shows one example where the drone is clearly visible at around 450~m distance from the \glspl{ru}. The noise-plus-interference-power has been measured in the areas marked by the black dotted lines. These areas in the periodogram were selected to minimize the contribution of clutter and just obtain a noise plus interference ’floor’ by selecting the median value of the periodogram power distribution in this area.

\begin{figure}
    \centering
    \includegraphics[width=0.95\linewidth]{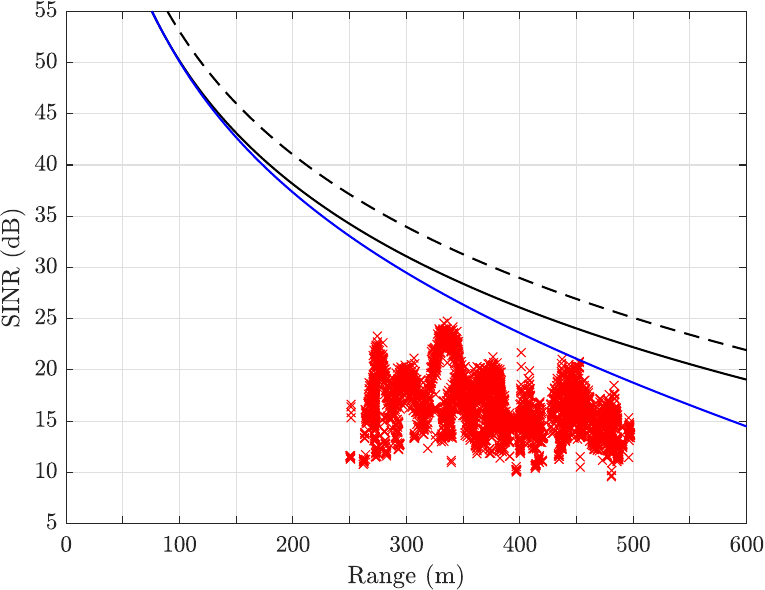}
    \caption{Measured target \gls{sinr} vs. range from \gls{isac} for the second experiment ({\color[rgb]{1,0,0}{\fbseries$\times$}}); expected SNR assuming thermal noise only ({\color[rgb]{0,0,0}\rule[0.5ex]{0.3em}{.7pt}\hspace{0.3em}\rule[0.5ex]{0.3em}{.7pt}}); expected SINR based on model in (1) ({\color[rgb]{0,0,0}\rule[0.5ex]{1em}{.7pt}}); expected SINR with additional consideration of FFT window length in (9) ({\color[rgb]{0,0,1}\rule[0.5ex]{1em}{.7pt}})}
    \label{fig:exp2_sinr}
\end{figure}

Fig.~\ref{fig:exp2_sinr} shows the measured \gls{sinr} vs. range during the complete flight. Note that only the \gls{sinr} of detected peaks is shown, but not of those not detected, though anyway known from the ground truth position of the \gls{uav}. We observe that correctly detected and validated peaks in our experiment have an \gls{sinr} of at least $\approx 10$~dB. 
We further observe the fluctuating behavior of the \gls{sinr} which is caused by the signal power (and not by noise-plus-interference) and dominates in our experiment over the dependency from increasing and decreasing range. One possible explanation for this effect is the time variation of the \gls{rcs} $\sigma$ during the flight and is matter of further investigation. The expected \gls{sinr} in Fig.~\ref{fig:exp2_sinr} assumes full antenna gains $G_\Tx$ and $G_\Rx$ while the drone approaches the edge of the main beam at the near end of the trajectory at 250~m distance which causes an attenuation of around 6~dB of the received signal strength and is visible as gap between expectation and experiment. 

The key outcome of this experiment is that our \gls{isac} system is capable to detect a small \gls{uav} well beyond $r_\mr{CP}=89$~m in 500 meters distance in the presence of strong clutter while flying faster than 5~m/s. We compare this result with the theoretically expected maximum detectable distance based on our link budget model in Section~\ref{sec:lb_model}.
Applying the parameter values in Table~\ref{tab:linkbudgetparams} and (\ref{eq:maxrange}), we calculate a maximum detectable distance of $r_\mr{max}=540~\mr{m}$ (17~dB \gls{sinr}) with a sum coupling loss of $\mi{C}_\mr{total} = 85~\mr{dB}$ caused by a building in 30~m distance as dominant clutter object. Note that we didn't adjust the reception window according to the expected target range during the experiment. The absolute range limit with optimized reception window and \gls{ofdm} symbol transmission aligned as in communication is $r_\mr{limit}=1338$~m.

\begin{table}[t]
    \caption{Most important settings and results of experiments. \label{tab:accuracy}}
    \centering
    \begin{tabu}{|l|c|c|}
         \hline
          & \textbf{Experiment 1}  & \textbf{Experiment 2} \\
         \Xhline{3\arrayrulewidth}
         Investigated range &0-90~m  &250-500~m \\
         \hline
         Beam steering &sweep over 6 beams &one fixed beam \\
         \hline
         Best clutter removal &CRAP &ECA-C \\
         \hline
         CA-CFAR false alarm rate &$10^{-6}$ &$10^{-4}$ \\
         \hline
         50\%-quantile of range error &0.09~m &0.23~m \\
         \hline
         95\%-quantile of range error &0.30~m &0.73~m \\
         \hline
    \end{tabu}
\end{table}
\section{Conclusion}\label{sec:conclusion}

We have presented two experiments for \gls{uav} detection with our mono-static \gls{isac} \gls{poc} in a realistic outdoor scenario and compared the measured results with the theoretical bounds from a link budget model. The objective of the first experiment with a flight route closer to the \glspl{ru} was to show that reliable detection of a \gls{uav} is possible even during a beam sweep, when the target is intermittently outside of the main cone of the beam, or even completely outside of the coverage area. The median of the discrepancy of the distance measured with \gls{isac} and with GNSS as reference is 9 cm if both systems are perfectly calibrated and synchronized.

The objective of the second experiment was to show that the \gls{uav} is detectable at long distances in the presence of strong static clutter. This is possible at least up to 500~m, the upper limit of the flight route, which is close to the expected limit of around 540~m derived from the link budget model. This model considers constraints coming from the design of the hardware and the choice of system parameters tailored for pure communication services. Gaps in the detection during the flight are mainly caused by the time-variant behaviour of the \gls{rcs} that, in turn, leads to fluctuations of the \gls{sinr}, and too low radial speed smaller than 5~m/s in combination with the required suppression of static clutter in the periodogram The median discrepancy between measured range and the GNSS reference is 23~cm.

\balance
\section*{Acknowledgments}
The authors acknowledge the financial support by the Federal Ministry of Research, Technology and Space of Germany in the project SENSATION under grant number 16KIS2523K.

\bibliographystyle{IEEEtran}
\bibliography{drone}

\end{document}